\documentclass[a4paper,12pt]{article}

 \newcommand{\be}{\begin{equation}}
 \newcommand{\ee}{\end{equation}}

  \usepackage{amsmath,epsfig,amssymb,amsfonts,verbatim}
  \usepackage[english]{babel}
  \usepackage{hyperref}

 \hypersetup{colorlinks,%
 filecolor=magenta,%
 citecolor=cyan,%
 linkcolor=blue,%
 urlcolor=magenta,%
 pdftex}

\begin{document}

 \thispagestyle{empty}

  \begin{flushright}
  November 2006
  \end{flushright}

 \vspace{1cm}
 \begin{center}
 \Large{\bf  String Theory: A Theory of Unification}
 \end{center}

 \vspace{.5cm}
 \begin{center}
 {\sl Oswaldo Zapata}\\Email: \texttt{zapata.oswaldo@gmail.com}
 \end{center}

 \abstract{These notes on string theory are based on a series of talks I gave during my graduate studies.
 As the talks, this introductory essay is intended for young students and non-string theory physicists.}

 \section{Introduction}
 \label{int}\noindent

  For more than twenty years the standard model of particle physics has successfully
  described
 data from experiments probing  energies up to the order of 100 GeV.
 It has also predicted new phenomena and particles subsequently
 detected, and it continues to provide us with new insights into
 the microscopic physical world. In spite of the fact that the
 standard model
 is the most reliable theory we have to describe phenomena at
 subnuclear scales, at higher energies new physics is expected to
 take place and new theoretical frameworks are required. A viable
 model is string theory, a theory describing physics at very such
 high energies that quantum gravitational effects cannot be
 neglected (Planck scale $\sim$ 10$^{19}$ GeV). It is worth
 stressing that in general a new high energy description do not need to be a simple
 extension of the standard model, in fact, even if the former should reduce to it in
 the limit of low energies, it can be different at the
 conceptual level. This is the case of string theory, where
 new concepts such as worldsheet, branes, extra dimensions and so
 on are introduced. In this essay I shall focus on string theory,
 a possible quantum theory of gravity and moreover expected to be
 a unified model of all the fundamental interactions. In the last
 part I comment on a novel approach relating confining gauge
 theories and string theory: the AdS/CFT
 correspondence.

 \section{ Symmetries and Unification }
 \label{symm}\noindent

     As in many other branches of physics, symmetries also play a crucial role
 in the search for a unified theory of nature. In particular, a new
 symmetry relating bosons and fermions has emerged as a fundamental
 ingredient. This symmetry is called supersymmetry (SUSY) \cite{susy} and has
 changed our understanding of the spacetime in a dramatic way.
 Indeed, SUSY is the only possible extension of the
 four-dimensional Poincar\'e symmetry, and it is built adding
 anticommuting directions to the usual spacetime. The spectrum of
 the supersymmetric extension of the Poincar\'e algebra indicates
 that every boson has a fermionic supersymmetric partner and vice
 versa. At low energies there is no experimental evidence for
 supersymmetric particles, nevertheless, at high energies it is
 hoped that SUSY becomes an exact symmetry.

  In addition to this, SUSY is considered a fundamental principle
  of any unified theory. In a unified description of all the fundamental
 interactions it is required that the coupling constant of each
 force, the parameter characterizing its strength, coincides with
 all the others at the energy of unification. But, the couplings of the
 standard model in general depend on the particle content and it is known that
 they never coincide in a common value, nevertheless, including
 supersymmetry the spectrum gets modified in such a way that it
 provides a better possibility for unification (E $\sim$ 10$^{17}$
 GeV).

  The formulation of a viable grand unified theory (GUT) is one of
 the main motivations for studying SUSY, another reason is that
 SUSY gives valuable hints in order to solve the related hierarchy
 problem. The latter is a non-trivial puzzle of the standard model
 and it concerns the existence of two very different energy scales
 in the model. One of these is the electroweak scale ($\sim$ 100
 GeV) and the other one is a very high energy supposed to be the
 the Planck scale. The presence of such huge energy is
 theoretically suggested in the low energy dynamics by the well
 known divergent radiative corrections to scalar masses\footnote{This
 suggests that the standard model must be thought of
 as an effective field theory of a more fundamental theory incorporating
 gravity.}. Hence, to solve the
 hierarchy problem the divergences of the standard model should be
 smoothed. Introducing SUSY this is achieved since an equal number
 of propagating bosonic and fermionic degrees of freedom cancel
 each other in loop diagrams. With this simple artifact SUSY solves
 the technical aspects of the hierarchy problem.

  Another related long standing problem where SUSY has also played an
 important role is in the reconciliation of general relativity and
 quantum field theory, that is, in the formulation of a quantized
 theory of gravity. An elegant scheme for this is provided by
 supergravity (SUGRA) \cite{sugra}, a model where supersymmetry is defined
 locally. It turns out that the spectrum of the theory contains a
 particle whose dynamics is dictated by Einstein equations, so such
 particle is naturally identified with the graviton. However, the
 introduction of SUSY does not get rid of all the infinities
 this quantized theory of gravity shows. Actually, at short
 distances it can be shown that renormalizability becomes useless.
 This is a reliable sign that SUGRA, local super-Poincar\'e, is
 also a low energy regime of a more fundamental theory. Nowadays, a large
 amount of theoretical evidences converge to indicate that this
 formulation is realized by string theory.

 \section{ String Theory }
 \label{stringth}\noindent

   String theory \cite{superstrings} was originally proposed as a model,
   the old dual string model,
 for describing strong nuclear interactions. However, it was
 discarded due to the success of quantum chromodynamics \cite{qcd},
 the latter accounting satisfactorily for the great number
 of data arising at the time from high energy experiments (deep
 inelastic scattering). At that time, string theory suffered from two main
 physical deficiencies: first, the presence in its spectrum of a
 massless spin-2 particle not revealed in experiments and,
 secondly, its mathematical consistency only in spacetime
 dimensions larger than four.

  The first issue was solved noting that the presence of a massless
 spin-2 particle could indeed be evidence that the string model was
 something more than a simple theory of the strong nuclear force,
 in fact, after identifying the spin-2 particle with the graviton,
 the model was elevated to a quantized theory of gravity.
 Furthermore, very remarkably, it was shown that the low energy
 limit of string theory coincided with some already known SUGRA, where it is
 known that at lower energies the graviton interacts as in general
 relativity.

 The trouble caused by the extra dimensions was overcome
 borrowing an old idea from Kaluza and Klein, \emph{i.e.}, the extra
 dimensions are thought to be so small, more or less of the same
 size than the string length, that they cannot be perceived in
 daily life experience. More recently it was conjectured that the
 extra dimensions could even attain the TeV scale. These new models
 are collectively called brane-worlds and it is eagerly expected
 that they will give rise to testable results in high energy
 physics or even in cosmology.

 The crucial difference between strings and point particles
 is that strings have an infinite internal number of degrees of
 freedom, they can vibrate, while point particles can only
 propagate in space. This freedom to vibrate, with each oscillation
 associated with a particle, gives rise to a characteristic
 spectrum. In addition to the ten-dimensional graviton, the
 spectrum also contains the dilaton field $\phi$ whose vacuum
 expectation value determines the unique string coupling constant
 $g_s=e^{\langle\phi\rangle}$ and allows for a perturbative
 analysis of the theory.

 In contrast to supergravity string theory is free of
 ultraviolet divergences. Let us see how this works. Generalizing
 quantum field theory techniques, a perturbative formulation of
 string theory
 is carried out summing over all possible histories of stringy
 Feynman topological diagrams, each of them with certain power of
 the coupling. For example, each genus $h$ Riemann surface
 representing an interacting level of closed strings comes with a
 factor $g_s^{2h+2}$ with $h=0,1,2,\dots$ Proving that each of
 these stringy diagrams is ultraviolet finite the theory overcomes
 the problem with short distance singularities of supergravity and
 other point particle models.

  In addition to the one-dimensional fundamental string,
 superstring theories also include higher dimensional objects
 called D$p$-branes. These objects are $p$-dimensional black hole
 type solutions of SUGRA (see below) and are the natural higher
 dimensional generalization of point particles ($p=0$) and strings
 ($p=1$). Moreover, like point particles can be electrically
 charged, D$p$-branes can carry Ramond-Ramond charges. The $D$ is
 for Dirichlet and indicates that open strings have their end
 points living on the brane. Therefore, open strings are
 constrained to move with their ends attached to these hyperplanes.
 On the other hand, closed strings can move freely in all the
 space. This picture suggests that the only manner an open string
 can get away from the brane is if both ends join at some point,
 form a closed string and then escape far away. Since open string
 excitations are associated with gauge fields living on the branes,
 whereas gravitational fields are mediated by closed strings, whose
 lines of force can invade all the dimensions, we are lead to think
 that there is a closed connection between particle and
 gravitational physics.

 \section{ Nuclear Forces and Strings}
 \label{nuclear}\noindent

   At high energies non-Abelian gauge theories as QCD are weakly coupled,
 opening the possibility for exact predictions. On the other hand,
 in the infrared the theory is strongly coupled and a perturbative
 analysis is nonsense. The strong behavior QCD shows at low
 energies explain why quarks, the constituents of nucleons, cannot
 be isolated in experiments and remain coupled to each others
 forming different composite particles. Confinement is an important
 unsolved problem in particle physics that string theory has
 contributed to clarify.

 The old dual string model supposed that the chromo-electric
 fields emitted by quarks, and binding them together, were thin
 tubes or strings (the `QCD-string'). In this picture confinement finds a natural
 explanation since farer the quarks are brought apart greater is
 the energy needed to continue separating them. Nevertheless,
 strictly speaking QCD is not a confining theory since at energies
 beyond the QCD scale $\Lambda_{QCD}\sim$ 300 MeV the vacuum starts
 to create quark-antiquark pairs and thus the original quarks can
 be separated as much as desired. Since this last phenomenon is
 still more complex than pure confinement theorists have
 concentrated mostly on the latter.

  QCD-like theories, including their supersymmetric extensions,
 are real confining theories as long as they satisfy at least one
 of the two following conditions: all the quarks are much heavier
 than the QCD scale, the point where the vacuum breaks, or the
 number of family of quarks is very large. The importance of these
 variants is that confinement survives these limits and many
 simplifications occur, allowing for new theoretical insights of
 the real problem.

  The introduction of a large number of colors, $N \to \infty$,
 changes the expansion parameter of the original Yang-Mills theory
 $g_{YM}$ to a new effective 't Hooft Coupling $\lambda\equiv
 g_{YM}^2N$. The new expansion parameter now allows for a
 Feynman-like diagrammatic in Riemann surfaces with increasing
 genus. This is an old result and it is striking how similar it is
 to the perturbative expansion of string theory we saw above,
 nevertheless, only recently it was shown that large $N$ gauge
 theories and string quantum gravity have indeed an equivalent
 (dual) mathematical description.

 \section{ String / Gauge Theory Duality }
 \label{stringgauge}\noindent

  Maxwell equations with magnetic monopoles are invariant under the
 interchange of electric and magnetic quantities. The two theories
 are said to be dual. The classical comparison of these theories is
 not interesting since they are completely analogous, nevertheless,
 what is relevant is that the dual quantized versions of them are quite different. What in a
 first moment was a perturbative analysis in terms of the weak
 electric coupling, in the dual magnetic description becomes a
 non-perturbative strongly coupled system. The D-branes
 mentioned above, with their tension going as $g_s^{-1}$, are the
 analogue in string theory of these dual magnetic objects. In both
 cases they help to probe the physics at strong coupling. The
 AdS/CFT conjecture we shall study below makes use of this idea,
 now proposing a duality between a gravity and a non-Abelian gauge
 theory \cite{agmoo}.

  In order to clarify this duality it is convenient to introduce
  the concept of black hole,
  a very massive object originated in a gravitational collapse,
  inside of which all the forces of nature are in action.
  For our purposes, a black hole \cite{townsend} simply can be regarded as a region of
  spacetime from where no information can
  escape beyond its boundary,
  \emph{i.e.,} the information inside the black hole is inaccessible to
  distant observers. Moreover, black holes are very simple objects since
  their properties do not depend
  on the kinds of constituents they are made of,
  but instead on some basic properties such as mass, charge and
  angular momentum.

  The Schwarzschild black hole is the simplest one and it has
  a event horizon which is a sphere of area $A=4\pi G^2M^2/c^4$.
  It can be systematically proved that
  this area cannot decrease in any classical process.
  On the other hand, gravitational collapsing objects
  which give rise to black holes seem to violate the
  second law of thermodynamics. This is easy to see since the initial
  collapsing object has
  a non-vanishing entropy whereas the final black hole cannot radiate, then the
  entropy of the entire system has decreased. The problem is solved by
  providing an entropy to the black hole.
  For a Schwarzschild black hole it was proposed by Bekenstein that
  the entropy is proportional to the event horizon area,
  a quantity that can only increase as the entropy does in classical
  thermodynamics,
  \be
  S_{BH}=\frac{1}{4}\frac{A}{l_p^2}~.
  \ee

  The generalized second law of thermodynamics
  extends the usual second law to
  include the entropy of black holes in a composite system, counting the
  entropy of the standard matter system and also that of the black hole
  $S_{TOT}=S_{MATT}+S_{RAD}+S_{BH}$. This is the entropy that
  always increases.
  Starting with
  a collapsing object of entropy $S$, the generalized second law
  of thermodynamics imposes
  that $S\leq S_{BH}$.
  This is the holographic bound,
  and it states that the entropy of a matter system entirely
  contained inside a surface of area $A$,
  cannot exceed that of a black hole of the same size.
   Alternatively, the holographic bound can be rephrased saying that the information of
  a system is completely stored in its boundary surface.

    This statement is generalized by the holographic principle \cite{holpr}. It
  claims that any physical process occurring in
  $D+1$ spacetime dimensions,
  as described by a quantum theory of gravity, can be equivalently described by
  another theory, without gravity, defined on its $D$-dimensional boundary.
  Some authors believe that this statement is universal and
  a fundamental principle of nature.
  Nevertheless, the principle has been tested only in a few
  concrete cases. An exception is the AdS/CFT correspondence,
  since it exactly relates superstrings in
  a $D$-dimensional space with a superconformal field theory on
  the boundary.

   Finally, let us comment on the black hole
  information paradox and see how the
  holographic principle resolves it. The paradox can be posed
  in the following terms. If the initial collapsing object is in a pure
  quantum state before it starts to contract, we expect
  the final object to be in exactly the same configuration. However, the
  thermal radiation of the final object comes necessarily as mixed states and so the information
  we get about the inside does not
  reproduce the information booked in the original object. We can say
  that the initial information is lost or destroyed inside the black hole. This
  paradox is solved by the holographic principle since the
  full dynamics of the gravitational theory is now described by a
  standard, though complex, quantum system with unitary
  evolution.

  So far the most accurate holographic proposal relating
  gauge theories to a quantized theory of gravity is the novel AdS/CFT correspondence.
  In two words, it says that string theory defined in a negatively curved anti-de Sitter
  space (AdS) is equivalent to a certain conformal field theory (CFT)
  living on its boundary.
  One concrete example is AdS$_5$/CFT$_4$: it states that
  type IIB superstring theory in
  AdS$_5$ is
  equivalently described  by an extended ${\cal N}=4$ super-CFT
  in four dimensions. The other five
  dimensions of the bulk are
  compactified on S$^5$. The five-sphere with isometry group
  $SO(6)$ is chosen in order to match with
  the $SU(4)$ R-symmetry of the super Yang-Mills theory.

    The AdS$_5$/CFT$_4$ correspondence can be motivated as follows. Starting with $N$
 parallel D3-branes on top of each other, we have open strings
 ending on the branes and closed strings moving in the whole space.
 On the branes the open string dynamics is that of a weakly coupled
 gauge theory. Since D$p$-branes are black holes the latter
 description should coincide with that of strongly interacting
 closed strings near the horizon. Hence, the strongly coupled
 gravitational theory in the bulk is equivalent to a weakly coupled
 gauge theory on the brane, and viceversa. This is the basic
 statement of the AdS/CFT conjecture.

   Since the AdS/CFT correspondence is a weak/strong duality it allows us to probe the
 strong coupling regime of the gauge theory with a perturbative
 analysis in the string side. This statement is established by the fundamental relation between
 the two sides of the correspondence
  \be\label{adscftfund}
  \left( \frac{R}{l_s}  \right)^4= g_s N~\leftrightarrow~ g_{YM}^2N\equiv \lambda~,
  \ee
  where $R$ is the curvature radius of the anti-de Sitter space,
  $N$ is the number of colors (in the large limit $N\to \infty$) in the gauge group and $\lambda$ is
  the 't\,Hooft coupling.

     \qquad   In the supergravity limit the string length is
  much smaller than
  the radius of the AdS space, given
  \be
  1\ll  \left( \frac{R}{l_s}  \right)^4 ~\leftrightarrow~ \lambda~.
  \ee
  In this limit the bulk theory is manageable, being a gauged supergravity theory, but in the
  boundary side it turns out that the gauge theory is in a strongly coupled regime, where a
  perturbative analysis is senseless. This establishes the weak/strong coupling nature of the duality.
  This is a clear advantage is we want to
  study the strongly coupled regime of one of the theories, since we can
  always use perturbative results in the dual theory. However,
  the difficulties in finding a common perturbative sector where to
  test the correspondence makes it hard to prove its full validity.
  The strong formulation of the
  AdS/CFT correspondence claims its validity at the string quantum level,
  nevertheless,
  so far nobody has been able to quantitatively prove it beyond the
  supergravity approximation.

  The main challenges of AdS/CFT are two-fold: i)~ to shed light in the strongly coupled
 regime of non-Abelian gauge  theories, as a step further in the
 understanding of more realistic QCD-like theories; ii)~ to provide
 a full proof of the correspondence. The latter is a non-trivial
 task since we do not have an independent non-perturbative
 definition of string theory that could be compared with the
 boundary theory at the strong regime. Pointing in this direction,
 a couple of years ago a new proposal was suggested, that goes
 under the name of BMN conjecture \cite{bmn,jabbari}, and opened the possibility to
 test the correspondence beyond the SUGRA limit. The idea is to
 investigate the consequences certain limiting procedure, namely,
 the Penrose
 Limit, has on both sides of the correspondence. In this limit not
 all the ideas involved are conceptually well established. One of
 these is the fate of holography in the BMN limit. It seems that
 the beautiful holographic picture that the AdS/CFT duality shows
 is completely lost in the plane-wave background.\\
 \\
 {\bf Acknowledgements:} I am grateful to M. Bianchi, M. Borunda, S.
 Kovacs and B. Schroer for a critical reading of the manuscript.

 \end{document}